\documentclass[prd,aps,showpacs,tightenlines,nofootinbib,preprint,eqsecnum]
{revtex4}
\usepackage{amssymb}
\usepackage{amsmath}

\newcommand{\bear}{\begin{array}}  \newcommand{\eear}{\end{array}}
\newcommand{\bea}{\begin{eqnarray}}  \newcommand{\eea}{\end{eqnarray}}
\newcommand{\beq}{\begin{equation}}  \newcommand{\eeq}{\end{equation}}
\newcommand{\bef}{\begin{figure}}  \newcommand{\eef}{\end{figure}}
\newcommand{\bec}{\begin{center}}  \newcommand{\eec}{\end{center}}

\newcommand{\bib}{\bibitem}

\newcommand{\Eqn}[1]{&\hspace{-0.2em}#1\hspace{-0.2em}&}

\def\APJ#1#2#3{Astrophys. J. {\bf #1}, #2 (19#3)}
\def\APJJ#1#2#3{Astrophys. J. {\bf #1}, #2 (20#3)}

\def\ARAA#1#2#3{Ann. Rev. Astron. Astrophys. {\bf#1}, #2 (19#3)}
\def\ARAAA#1#2#3{Ann. Rev. Astron. Astrophys. {\bf#1}, #2 (20#3)}

\def\IJMPDD#1#2#3{Int. J. Mod. Phys. D {\bf #1}, #2 (20#3)}

\def\JHEPP#1#2#3{J. High Energy Phys. {\bf #1}, #2 (20#3)}

\def\MNRAS#1#2#3{Mon. Not. R. Astron. Soc. {\bf #1}, #2 (19#3)}
\def\MNRASS#1#2#3{Mon. Not. R. Astron. Soc. {\bf #1}, #2 (20#3)}
\def\MPLA#1#2#3{Mod. Phys. Lett. A {\bf #1}, #2 (19#3)}

\def\PLB#1#2#3{Phys. Lett. B {\bf #1}, #2 (19#3)}
\def\PLBB#1#2#3{Phys. Lett. B {\bf #1}, #2 (20#3)}
\def\PL#1#2#3{Phys. Lett. {\bf #1}, #2 (19#3)}

\def\PR#1#2#3{Phys. Rev. {\bf #1}, #2 (19#3)}

\def\PRD#1#2#3{Phys. Rev. D {\bf #1}, #2 (19#3)}
\def\PRDD#1#2#3{Phys. Rev. D {\bf #1}, #2 (20#3)}

\def\PRL#1#2#3{Phys. Rev. Lett. {\bf#1}, #2 (19#3)}
\def\PRLL#1#2#3{Phys. Rev. Lett. {\bf#1}, #2 (20#3)}

\def\PRTT#1#2#3{Phys. Rep. {\bf#1}, #2 (20#3)}

\def\PTPP#1#2#3{Prog. Theor. Phys. {\bf #1}, #2 (20#3)}

\def\RPP#1#2#3{Rep. Prog. Phys. {\bf #1}, #2 (19#3)}

\def\RMPP#1#2#3{Rev. Mod. Phys. {\bf #1}, #2 (20#3)}

\def\Vec#1{\mbox{\boldmath $#1$}}
\def\Vecs#1{\mbox{\boldmath\tiny $#1$}}

\def\Lap{{\mathop{\Delta}\limits^{(3)}}}

\begin{document}

\title{
The interrelation between the generation of large-scale electric fields 
and that of large-scale magnetic fields during inflation
}

\author{Kazuharu Bamba}
\affiliation{
Department of Physics, Kinki University, 
Higashi-Osaka, Osaka 577-8502, Japan
}


\begin{abstract}
The interrelation between the generation of large-scale electric fields 
and that of large-scale magnetic fields due to the breaking of the conformal 
invariance of the electromagnetic field in inflationary cosmology is studied. 
It is shown that if large-scale magnetic fields with a sufficiently large 
amplitude are generated during inflation, the generation of 
large-scale electric fields is suppressed, and vice versa. 
Furthermore, a physical interpretation of the result and its 
cosmological significance are considered.
\end{abstract}


\hspace{13.0cm} KU-TP\ 015

\maketitle

\section{Introduction}

Magnetic fields with the field strength $\sim 10^{-6}$G on a $1-10$kpc scale 
have been observed in galaxies of all types, galaxies at cosmological 
distances \cite{Kronberg2}, and clusters of galaxies (for detailed reviews, 
see \cite{Sofue, Kronberg1, Grasso, Carilli1, Widrow, Giovannini1, Semikoz1}). 
Moreover, the field strength of magnetic fields in clusters of galaxies 
is estimated at $10^{-7}-10^{-6}$G and 
the scale is estimated at 10kpc$-$1Mpc scale~\cite{Kim1}. 
The origin of these magnetic fields, in particular 
magnetic fields in clusters of galaxies on a coherence scale 
as large as $\sim$Mpc, is not well understood yet. 
Although galactic dynamo mechanisms \cite{EParker} have been proposed 
to amplify very weak seed magnetic fields up to $\sim 10^{-6}$G, 
they require initial seed magnetic fields to feed on.  
Furthermore, the effectiveness of the dynamo amplification mechanism 
in galaxies at high redshifts or 
clusters of galaxies is not well established. 

There exist two broad categories of proposed generation mechanisms of seed 
magnetic fields. 
One is astrophysical processes~\cite{Biermann1, PI}, 
and the other is cosmological processes in the early universe, 
e.g., cosmological phase transition~\cite{Baym, Durrer, Quashnock, 
Boyanovsky1}, the generation of the magnetic fields from primordial 
density perturbations before the epoch of recombination 
\cite{Notari1, Ichiki1, Gopal1, Siegel1, Berezhiani1, Kobayashi07}. 
In fact, however, it is difficult that these processes generate 
the magnetic fields on megaparsec scales with sufficient field strength 
to account for the observed magnetic fields in galaxies and clusters of 
galaxies without requiring any dynamo amplification.  

The most natural origin of such a large-scale magnetic field is 
electromagnetic quantum fluctuations generated in the inflationary stage 
\cite{Turner}.  
This is because inflation naturally produces effects on very large scales, 
larger than Hubble horizon, starting from microphysical processes 
operating on a causally connected volume.  
Since the Friedmann-Robertson-Walker (FRW) metric 
usually considered is conformally flat and 
the classical electrodynamics is conformally invariant, 
the conformal invariance of the Maxwell theory must have been 
broken in the inflationary stage\footnote{
In Ref.~\cite{Maroto1}, the breaking of conformal flatness of the 
FRW metric induced by the evolution of scalar metric perturbations at 
the end of inflation has been discussed.} 
in order that electromagnetic quantum fluctuations could be generated 
at that time \cite{Parker}.  
Hence various conformal symmetry breaking mechanisms have been studied 
\cite{Turner, RF^2, Ratra, Scalar, Bamba1, Bamba2, Akhtari-Zavareh07, 
Charged-Scalar, ScalarED, Amplification, 
Dolgov1, Bertolami1, Gasperini1, Prokopec1, Enqvist1, 
Bertolami2, Ashoorioon1, Salim07, Giovannini:astro-ph/0612378, Bamba4}. 

It follows from indications in higher-dimensional theories including string 
theory that there can exist the dilaton field coupled to the electromagnetic 
field. 
Moreover, there can exist non-minimal gravitational couplings between 
the scalar curvature and the electromagnetic field due to one-loop 
vacuum-polarization effects in curved spacetime~\cite{Drummond}. 
These couplings break the conformal invariance of the electromagnetic field. 
Such a coupling of non-trivial background fields that vary in time to 
the electromagnetic field is very interesting as the generation mechanism of 
large-scale magnetic fields with a sufficiently large amplitude. 
Recently, therefore, in Ref.~\cite{Bamba3}, 
the present author and Sasaki studied the evolution of the 
electromagnetic field in a very general situation in which the conformal 
invariance is broken through the coupling of the form 
$I F_{\mu\nu}F^{\mu\nu}$ where $I$ can be a function of any non-trivial 
background fields that vary in time,
and $F_{\mu\nu} = {\partial}_{\mu}A_{\nu} - {\partial}_{\nu}A_{\mu}$
is the electromagnetic field-strength tensor. Here, $A_{\mu}$ is 
the $U(1)$ gauge field. 
In this case, not only large-scale magnetic fields but also large-scale 
electric fields can be generated during inflation. 
The conductivity of the universe in the inflationary stage 
is negligibly small, because there are few charged particles at that time. 
Hence electric fields can exist during inflation. 

In the present paper we consider the interrelation between the generation of 
large-scale electric fields and that of large-scale magnetic fields during 
inflation due to the breaking of the conformal invariance of the 
electromagnetic field through a coupling with non-trivial background fields 
that vary in time. 
We show that when large-scale magnetic fields with a sufficiently large 
amplitude are generated during inflation, the generation of 
large-scale electric fields is suppressed, and vice versa. 
In the inflationary stage, the sum of the energy density of electric and 
magnetic fields should be smaller than that of the inflaton field. 
From our result, we find that 
there does not exist the possibility that 
if large-scale magnetic fields with a sufficiently large 
amplitude are generated during inflation, 
large-scale electric fields with a sufficiently large 
amplitude are generated simultaneously, so that 
the sum of the resultant energy density of electric and magnetic 
fields becomes larger than that of the inflaton. 
If large-scale electric fields with a large amplitude are generated, 
the large-scale charge separation and additional fluctuations 
in the cosmic plasma could be generated during reheating, so that 
the evolution of the universe might become anisotropic. 
In fact, however, in this scenario when the large-scale magnetic fields with 
a sufficiently large amplitude are generated during inflation, 
the amplitude of the generated large-scale electric fields is very small. 
Hence the large-scale charge separation and additional 
fluctuations in the cosmic plasma can be hardly generated. 
Thus, this generation scenario of large-scale magnetic fields from 
inflation is consistent with the standard evolution of the universe 
suggested from the observation of the cosmic microwave background (CMB) 
radiation. 
We use units in which $k_\mathrm{B} = c = \hbar = 1$. 
Moreover, we adopt Heaviside-Lorentz units in terms of electromagnetism. 

This paper is organized as follows. 
In Sec.\ II we review the model in Ref.~\cite{Bamba3} and consider 
the evolution of the $U(1)$ gauge field. 
In Sec.\ III we derive the energy density of large-scale electric and 
magnetic fields generated during inflation, and we consider the interrelation 
between the generation of large-scale electric fields and 
that of large-scale magnetic fields. 
Furthermore, we consider a physical interpretation of the result and its 
cosmological significance in Sec.\ IV. 
Finally, Sec.\ V is devoted to a conclusion.

\section{Breaking of the conformal invariance of the electromagnetic field}

\subsection{Model}

First, we review the model in Ref.~\cite{Bamba3}. 
We consider the following model action: 
\begin{eqnarray}
S = 
\int d^{4}x \sqrt{-g} 
\left(-\frac{1}{4}I\,
F_{\mu\nu}F^{\mu\nu}\right)\,,
\label{eq:2.1} 
\end{eqnarray} 
where $g$ is the determinant of the metric tensor $g_{\mu\nu}$, 
and $I$ is an arbitrary function of non-trivial background fields 
at the moment. 

{}From the action~(\ref{eq:2.1}), 
the equation of motion for the electromagnetic field can be derived as 
follows: 
\begin{eqnarray}
-\frac{1}{\sqrt{-g}}{\partial}_{\mu} 
\left[ \sqrt{-g} I F^{\mu\nu} 
\right] = 0.  
\label{eq:2.2}  
\end{eqnarray}

We assume the spatially flat FRW space-time with the metric 
\begin{eqnarray}
{ds}^2 =-{dt}^2 + a^2(t)d{\Vec{x}}^2
= a^2(\eta) ( -{d \eta}^2 + d{\Vec{x}}^2 ), 
\label{eq:2.3}
\end{eqnarray} 
where $a$ is the scale factor, and $\eta$ is the conformal time.  
We consider the evolution of the $U(1)$ gauge field in this background.  
Its equation of motion in the Coulomb gauge, 
$A_0(t,\Vec{x}) = 0$ and ${\partial}^jA_j(t,\Vec{x}) =0$, reads 
\begin{eqnarray}
\ddot{A_i}(t,\Vec{x}) 
+ \left( H + \frac{\dot{I}}{I} 
\right) \dot{A_i}(t,\Vec{x}) 
- \frac{1}{a^2}\Lap\, A_i(t,\Vec{x}) = 0\,,
\label{eq:2.4}
\end{eqnarray}
where $H=\dot a/a$ is the Hubble parameter, and a dot denotes 
a time derivative, $\dot{~}=\partial/\partial t$. Moreover, 
$\Lap =  {\partial}^i {\partial}_i$ is the flat 3-dimensional 
Laplacian. 

\subsection{Evolution of the $U(1)$ gauge field}

Next, we consider the evolution of the $U(1)$ gauge field 
in generic slow-roll inflation. 
Here we shall quantize the $U(1)$ gauge field $A_{\mu}(t,\Vec{x})$. 
It follows from the model Lagrangian in Eq.~(\ref{eq:2.1}) that the canonical 
momenta conjugate to $A_{\mu}(t,\Vec{x})$ are given by 
\begin{eqnarray}
{\pi}_0 = 0, \hspace{5mm} {\pi}_{i} = I a(t) \dot{A_i}(t,\Vec{x}).
\label{eq:2.5} 
\end{eqnarray}
We impose the canonical commutation relation 
between $A_i(t,\Vec{x})$ and ${\pi}_{j}(t,\Vec{x})$, 
\begin{eqnarray} 
  \left[ \hspace{0.5mm} A_i(t,\Vec{x}), {\pi}_{j}(t,\Vec{y}) 
  \hspace{0.5mm} \right] = i
 \int \frac{d^3 k}{{(2\pi)}^{3}}
             e^{i \Vecs{k} \cdot \left( \Vecs{x} - \Vecs{y} \right)}
        \left( {\delta}_{ij} - \frac{k_i k_j}{k^2 } \right),
\label{eq:2.6} 
\end{eqnarray}
where $\Vec{k}$ is comoving wave number and $k=|\Vec{k}|$.  
{}From this relation, we obtain the expression for $A_i(t,\Vec{x})$ as 
\begin{eqnarray} 
\hspace{-4mm} A_i(t,\Vec{x}) = \int \frac{d^3 k}{{(2\pi)}^{3/2}}
  \sum_{\sigma=1,2}\left[ \hspace{0.5mm} \hat{b}(\Vec{k},\sigma) 
        \epsilon_i(\Vec{k},\sigma)A(t,k)e^{i \Vecs{k} \cdot \Vecs{x} }
       + {\hat{b}}^{\dagger}(\Vec{k},\sigma)
       \epsilon_i^*(\Vec{k},\sigma)
         {A^*}(t,k)e^{-i \Vecs{k} \cdot \Vecs{x}} \hspace{0.5mm} \right],
\label{eq:2.7} 
\end{eqnarray}
where $\epsilon_i(\Vec{k},\sigma)$ ($\sigma=1,2$)
are the two orthonormal transverse polarization vectors, 
and $\hat{b}(\Vec{k},\sigma)$ and 
${\hat{b}}^{\dagger}(\Vec{k},\sigma)$ are the annihilation and creation 
operators which satisfy 
\begin{eqnarray} 
\hspace{-5mm}  \left[ \hspace{0.5mm} \hat{b}(\Vec{k},\sigma),
 {\hat{b}}^{\dagger}({\Vec{k}}^{\prime},\sigma') \hspace{0.5mm} \right] = 
\delta_{\sigma,\sigma'}
{\delta}^3 (\Vec{k}-{\Vec{k}}^{\prime}), \hspace{3mm}
\left[ \hspace{0.5mm} \hat{b}(\Vec{k},\sigma),
 \hat{b}({\Vec{k}}^{\prime},\sigma')
\hspace{0.5mm} \right] = 
\left[ \hspace{0.5mm} 
{\hat{b}}^{\dagger}(\Vec{k},\sigma),
 {\hat{b}}^{\dagger}({\Vec{k}}^{\prime},\sigma')
\hspace{0.5mm} \right] = 0.
\label{eq:2.8} 
\end{eqnarray}
It follows from Eq.~(\ref{eq:2.4}) that the mode function $A(k,t)$ 
satisfies the equation 
\begin{eqnarray} 
\ddot{A}(k,t) + \left( H + \frac{\dot{I}}{I} \right) 
               \dot{A}(k,t) + \frac{k^2}{a^2} A(k,t) = 0, 
\label{eq:2.9}  
\end{eqnarray} 
and that the normalization condition for $A(k,t)$ reads 
\begin{eqnarray} 
A(k,t){\dot{A}}^{*}(k,t) - {\dot{A}}(k,t){A^{*}}(k,t)
= \frac{i}{I a}\,.
\label{eq:2.10} 
\end{eqnarray}
Replacing the independent variable $t$ by $\eta$, we find that 
Eq.~(\ref{eq:2.9}) becomes 
\begin{eqnarray}
A^{\prime \prime}(k,\eta) + 
\frac{I^{\prime}}{I} A^{\prime}(k,\eta) 
+ k^2 {A}(k,\eta) = 0,  
\label{eq:2.11}
\end{eqnarray}
where the prime denotes differentiation with respect to the conformal time 
$\eta$. 

We are not able to obtain the exact solution of Eq.~(\ref{eq:2.11})
for the case in which $I$ is given by a general function of $\eta$. 
In fact, however, we can obtain an approximate solution with sufficient 
accuracy by using the Wentzel-Kramers-Brillouin (WKB) approximation on 
subhorizon scales and the long-wavelength approximation on superhorizon 
scales, and matching these solutions at the horizon crossing~\cite{Bamba3}. 

In the exact de Sitter background, we find 
$-k \eta = k/(aH)$, where $H$ is the de Sitter Hubble parameter. 
Moreover, at the horizon-crossing, $H=k/a$ is satisfied, and 
hence $-k \eta_k =1$ is satisfied. Here, $\eta_k$ is the conformal time 
at the horizon-crossing. 
The subhorizon (superhorizon) scale corresponds
to the region $k|\eta|\gg1$ ($k|\eta|\ll1$). 
This is expected to be also a sufficiently good definition for
the horizon crossing for general slow-roll, almost exponential 
inflation. 

The WKB subhorizon solution is given by
\begin{eqnarray}
A_{\mathrm{in}} (k,\eta) = 
\frac{1}{\sqrt{2k}} I^{-1/2} e^{-ik\eta}, 
\label{eq:2.12} 
\end{eqnarray} 
where we have assumed that the vacuum in the short-wavelength 
limit is the standard Minkowski vacuum. 

On the other hand, the solution on superhorizon scales, 
$A_{\mathrm{out}} (k,\eta)$, can be obtained 
by using the long-wavelength expansion in terms of $k^2$ and 
matching this solution with the WKB subhorizon solution in 
Eq.~(\ref{eq:2.12}) at the horizon crossing. 
The lowest order approximate solution of $A_{\mathrm{out}} (k,\eta)$ is 
given by~\cite{Bamba3}
\begin{eqnarray} 
A_{\mathrm{out}} (k,\eta) = 
C(k) + D(k) \int_{\eta}^{{\eta}_{\mathrm{R}}} 
\frac{1}{I \left( \Tilde{\eta} \right)} 
d \Tilde{\eta}, 
\label{eq:2.13} 
\end{eqnarray} 
where 
\begin{eqnarray}  
C(k) \Eqn{=} 
\left.
\frac{1}{\sqrt{2k}} I^{-1/2}  
\left[
1- \left( \frac{1}{2} I^{\prime} + i k I \right) 
\int_{\eta}^{{\eta}_{\mathrm{R}}} 
\frac{1}{I \left(\Tilde{\Tilde{\eta}}\right)} 
d \Tilde{\Tilde{\eta}} \right] e^{-ik\eta} 
\right|_{\eta = \eta_k}, 
\label{eq:2.14} \\[3mm] 
D(k) \Eqn{=} 
\left.
\frac{1}{\sqrt{2k}} I^{-1/2}  
\left( \frac{1}{2} I^{\prime} + i k I \right) 
e^{-ik\eta} 
\right|_{\eta = \eta_k}.
\label{eq:2.15} 
\end{eqnarray}  
Neglecting the decaying mode solution, 
from Eqs.~(\ref{eq:2.13}) and (\ref{eq:2.14}) we find that 
$|A(k,\eta)|^2$ at late times is given by 
\begin{eqnarray}
\left|A(k,\eta)\right|^2 
= |C(k)|^2 
= \frac{1}{2kI(\eta_k)}
\left|1- \left(\frac{1}{2}\frac{I^{\prime}(\eta_k)}{kI(\eta_k)} 
+ i\right)
e^{-ik\eta_k}k\int_{\eta_k}^{{\eta}_{\mathrm{R}}}
\frac{I(\eta_k)}{I \left(\Tilde{\Tilde{\eta}} \right)}
d\Tilde{\Tilde{\eta}}\,\right|^2\,,
\label{eq:2.16}
\end{eqnarray}
where ${\eta}_{\mathrm{R}}$ is the conformal time at the time of reheating 
after inflation.

\section{Evolution of large-scale electric and magnetic fields}

In this section, we consider the evolution of large-scale electric and 
magnetic fields. The proper electric and magnetic fields are given by 
\begin{eqnarray}
{E_i}^{\mathrm{proper}}(t,\Vec{x}) 
\Eqn{=} a^{-1}E_i(t,\Vec{x}) = -a^{-1}\dot{A_i}(t,\Vec{x}), 
\label{eq:3.1} \\[3mm]
{B_i}^{\mathrm{proper}}(t,\Vec{x})
\Eqn{=} a^{-1}B_i(t,\Vec{x}) = a^{-2}{\epsilon}_{ijk}{\partial}_j 
A_k(t,\Vec{x}),
\label{eq:3.2}    
\end{eqnarray} 
where $E_i(t,\Vec{x})$ and $B_i(t,\Vec{x})$ are the comoving electric and 
magnetic fields, and ${\epsilon}_{ijk}$ is the totally antisymmetric tensor
(\hspace{0.5mm}${\epsilon}_{123}=1$\hspace{0.5mm}).  

Using Eqs.~(\ref{eq:2.13}) and (\ref{eq:3.1}), we find 
\begin{eqnarray}
|{E}^{\mathrm{proper}}(k,\eta)|^2  
=2 \frac{1}{a^4} |A^{\prime}(k,\eta)|^2
=2 \frac{1}{a^4} \frac{|D(k)|^2}{|I(\eta)|^2}\,,
\label{eq:3.3}
\end{eqnarray}  
where the factor 2 comes from the two polarization degrees of freedom. 
The energy density of the large-scale electric fields 
in Fourier space is given by 
${\rho}_E(k,\eta) = (1/2) |{E}^{\mathrm{proper}}(k,\eta)|^2 I(\eta)$. 
Hence, 
multiplying ${\rho}_E(k,\eta)$ by the phase-space density, 
$4 \pi k^3/(2\pi)^3$, and using Eq.~(\ref{eq:3.3}), 
we obtain the energy density of the large-scale electric fields 
in the position space 
\begin{eqnarray} 
\rho_E (L,\eta) = 
\frac{1}{2} 
\frac{4\pi k^3}{(2\pi)^3}|{E}^{\mathrm{proper}}(k,\eta)|^2 I(\eta) 
=\frac{|D(k)|^2}{2 \pi^2 k} \frac{k^4}{a^4} \frac{1}{I(\eta)}, 
\label{eq:3.4}
\end{eqnarray}
on a comoving scale $L=2\pi/k$. 

Similarly, using Eqs.~(\ref{eq:2.16}) and (\ref{eq:3.2}), we find 
\begin{eqnarray}
|{B}^{\mathrm{proper}}(k,\eta)|^2  
=2\frac{k^2}{a^4}|A(k,\eta)|^2
=2\frac{k^2}{a^4}|C(k)|^2\,,
\label{eq:3.5}
\end{eqnarray}  
where the factor 2 comes from the two polarization degrees of freedom. 
The energy density of the large-scale magnetic fields 
in Fourier space is given by 
${\rho}_B(k,\eta) = (1/2) |{B}^{\mathrm{proper}}(k,\eta)|^2 I(\eta)$. 
Hence, 
multiplying ${\rho}_B(k,\eta)$ by the phase-space density, 
$4 \pi k^3/(2\pi)^3$, and using Eq.~(\ref{eq:3.5}), 
we obtain the energy density of the large-scale magnetic fields 
in the position space 
\begin{eqnarray}
\rho_B(L,\eta) = 
\frac{1}{2} 
\frac{4\pi k^3}{(2\pi)^3}|{B}^{\mathrm{proper}}(k,\eta)|^2 I(\eta) 
= \frac{k|C(k)|^2}{2\pi^2}\frac{k^4}{a^4} I(\eta), 
\label{eq:3.6}
\end{eqnarray}
on a comoving scale $L=2\pi/k$. 

In order to study the property of generation of large-scale electric and 
magnetic fields more clearly, we consider the case in which the 
coupling function of non-trivial background fields 
to the electromagnetic field, $I$, is given by a specific form as follows: 
\begin{eqnarray}
I(\eta)=I_{\mathrm{s}} 
\left( \frac{\eta}{\eta_{\mathrm{s}}} \right)^{-\alpha}\,,
\label{eq:3.7} 
\end{eqnarray}
where $\eta_{\mathrm{s}}$ is some fiducial time during inflation, 
$I_{\mathrm{s}}$ is the value of $I(\eta)$ at $\eta=\eta_{\mathrm{s}}$, 
and $\alpha$ is a constant. 
In this case, from Eqs.~(\ref{eq:2.15}) and (\ref{eq:3.7}) we find 
\begin{eqnarray} 
\frac{|D(k)|^2}{k} = \frac{\alpha^2 + 4}{8} I(\eta_k), 
\label{eq:3.8} 
\end{eqnarray}
where we have used $-k \eta_k =1$.  
Moreover, from Eqs.~(\ref{eq:2.14}) and (\ref{eq:3.7}) we find~\cite{Bamba3}
\begin{eqnarray}
k|C|^2=\frac{1}{2I(\eta_k)}\left|1-\frac{\alpha+2i}{2(\alpha+1)}
e^{-ik\eta_k}\right|^2 \equiv\frac{{\cal C}}{2I(\eta_k)}\,,
\label{eq:3.9}
\end{eqnarray}
where $-k \eta_k =1$ and ${\eta}_{\mathrm{R}} \approx 0$ have been used. 
Here, ${\cal C}$ is a constant of order unity. 
Thus, from Eqs.~(\ref{eq:3.4}) and (\ref{eq:3.8}) we obtain 
\begin{eqnarray} 
\rho_E (L,\eta) = \frac{\alpha^2 + 4}{16 \pi^2} \frac{k^4}{a^4} 
\frac{I(\eta_k)}{I(\eta)} 
= \frac{\alpha^2 + 4}{16 \pi^2} \left( \frac{k}{aH} \right)^{4+\alpha} H^4, 
\label{eq:3.10}
\end{eqnarray}
where $H$ is the Hubble parameter at the inflationary stage. 
In deriving the second equality in Eq.~(\ref{eq:3.10}), we have used 
the following relation: 
\begin{eqnarray} 
\frac{I(\eta_k)}{I(\eta)} = \left( \frac{\eta}{\eta_k} \right)^{\alpha} 
= \left( \frac{-k\eta}{-k\eta_k} \right)^{\alpha} 
= \left( \frac{k}{aH} \right)^{\alpha}, 
\label{eq:3.11}
\end{eqnarray}
where in deriving the last equality we have used $-k \eta_k =1$ and 
$-k\eta = k/(aH)$. 
Similarly, from Eqs.~(\ref{eq:3.6}) and (\ref{eq:3.9}) we obtain 
\begin{eqnarray} 
\rho_B (L,\eta) = \frac{{\cal C}}{4 \pi^2} \frac{k^4}{a^4} 
\frac{I(\eta)}{I(\eta_k)} = 
\frac{{\cal C}}{4 \pi^2} \left( \frac{k}{aH} \right)^{4-\alpha} H^4, 
\label{eq:3.12}
\end{eqnarray}
where in deriving the second equality we have used 
the relation ~(\ref{eq:3.11}). 
Here we note that since we are interested in large-scale electric and 
magnetic fields, we consider the superhorizon scale, i.e., $k/(aH) \ll 1$. 

Consequently, it follows from Eqs.~(\ref{eq:3.10}) and (\ref{eq:3.12}) that 
the ratio of the energy density of the large-scale electric fields to 
that of the large-scale magnetic fields is given by 
\begin{eqnarray} 
\frac{\rho_E (L,\eta)}{\rho_B (L,\eta)} = 
\frac{\alpha^2 + 4}{4{\cal C}} \left( \frac{k}{aH} \right)^{2\alpha}.
\label{eq:3.13}
\end{eqnarray}
Since we here consider the superhorizon scale, $k/(aH) \ll 1$, from 
Eq.~(\ref{eq:3.13}) we see that if $\alpha > 0$, 
$\rho_B (L,\eta) > \rho_E (L,\eta)$, and that if $\alpha < 0$, 
$\rho_B (L,\eta) < \rho_E (L,\eta)$.  
Hence, 
if large-scale magnetic fields with a sufficiently large 
root-mean-square (rms) amplitude are generated during inflation, 
the generation of large-scale electric fields is suppressed, and vice versa. 
This result holds true for the case in which $I$ is given by 
an arbitrary function of non-trivial background fields. From 
Eqs.~(\ref{eq:3.4}) and (\ref{eq:3.6}), we see that 
$\rho_E (L,\eta) \propto 1/I (\eta)$ and 
$\rho_B (L,\eta) \propto I (\eta)$. 
If large-scale magnetic fields with a sufficiently large 
amplitude are generated during inflation, 
the value of the coupling function $I$ must be extremely small 
in the beginning and increase rapidly over time during 
inflation~\cite{Bamba3}. 
In such a case, from the above relations 
we see that the generation of large-scale electric fields is suppressed. 

During inflation the sum of the energy density of electric and magnetic 
fields should be smaller than that of the inflaton. 
From Eq.~(\ref{eq:3.12}), we see that 
if $\alpha \approx 4$, the spectrum of large-scale magnetic fields is nearly 
scale-invariant; in this case the amplitude of generated large-scale magnetic 
fields at the present time can be sufficiently large to explain the 
observations of magnetic fields in galaxies and clusters of 
galaxies~\cite{Bamba1, Bamba3}. 
In this case, from the above consideration, during inflation 
the energy density of large-scale electric fields is much smaller than 
that of large-scale magnetic fields. 
Thus there does not exist the possibility that 
when large-scale magnetic fields with a sufficiently large 
amplitude are generated during inflation, 
large-scale electric fields with a sufficiently large 
amplitude are also generated, so that 
the sum of the energy density of electric and magnetic 
fields becomes larger than that of the inflaton.

\section{Physical interpretation and cosmological significance}

In this section, we consider a physical interpretation of the result 
in the preceding section, i.e., if large-scale magnetic fields with a 
sufficiently large amplitude are generated during inflation, 
the generation of large-scale electric fields is suppressed, and vice versa. 
Furthermore, we consider the cosmological significance of this result.

\subsection{Physical interpretation}

It follows from Eqs.~(\ref{eq:3.1}) and (\ref{eq:3.2}) that 
if the amplitude of the $U(1)$ gauge field $A_i(t,\Vec{x})$ varies in time, 
electric fields are generated; on the other hand, if the amplitude of 
$A_i(t,\Vec{x})$ varies in terms of space coordinates, magnetic fields are 
generated. 
When the amplitude of $A_i(t,\Vec{x})$ greatly varies in time, 
the relative difference of the amplitude of $A_i(t,\Vec{x})$ 
at each of the space-coordinate points 
becomes very small because the amplitude of $A_i(t,\Vec{x})$ greatly 
grows (or decays) at all the space-coordinate points equally. 
It follows from the relation 
$A^{\prime}(k,\eta) = -D(k)/I(\eta)$, 
which is derived from Eq.~(\ref{eq:2.13}), 
that this situation is realized if the value of $I$ decreases rapidly 
in time during inflation. 
On the other hand, when there can exist the large relative difference of 
the amplitude of $A_i(t,\Vec{x})$ at each of the space-coordinate points, 
the variation of the amplitude of $A_i(t,\Vec{x})$ in time 
must be small. This is because the relative difference of the amplitude of 
$A_i(t,\Vec{x})$ at each of the space-coordinate points is dissipated by 
the large variation of the amplitude of $A_i(t,\Vec{x})$ in time. 
This situation is realized if the value of $I$ increases rapidly in time 
during inflation. 
Thus, increasing $I$ which favors the small variation of 
the amplitude of $A_i(t,\Vec{x})$ in time leads to 
stronger magnetic fields and vice versa. 

Consequently, in this scenario there does not exist the possibility that 
both large-scale electric and magnetic fields with a sufficiently large 
amplitude are generated simultaneously. 
Hence large-scale magnetic fields with a sufficiently large 
amplitude can be generated during inflation without being inconsistent with 
the fact that the sum of the energy density of the generated electric and 
magnetic fields during inflation should be smaller than that of the inflaton. 
This point is the first cosmological significance of the result of 
the present paper.

\subsection{Cosmological consequences of the electric and magnetic fields 
after inflation}

Furthermore, we consider the cosmological consequences of the electric and 
magnetic fields after inflation. 
The conductivity of the universe in the inflationary stage 
is negligibly small, because there are few charged particles at that time. 
Hence the electric fields can exist during inflation. 
After reheating following inflation, however, a number of charged particles 
are produced, so that the conductivity of the universe immediately becomes 
much larger than the Hubble parameter at that time~\cite{Turner}. 
Hence the electric fields accelerate charged particles and finally dissipate, 
and only the magnetic fields can survive up to the present time. 

When there exist the electric fields during reheating, 
since the electric fields would be constant over horizon size patches, 
this would lead to the flow of a current. 
Furthermore, if 
$E L_H = E/H = \sqrt{6/(8\pi)} M_{\mathrm{Pl}} \sqrt{\rho_E/\rho} 
> 2 m_e$, 
which means $\rho_E/\rho > \left( m_e/ M_{\mathrm{Pl}} \right)^2$, 
the electric fields are strong enough to `short circuit' by 
electron/positron production. 
Here, $L_H$ is the horizon size, $M_{\mathrm{Pl}}$ is the Planck mass, 
$\rho$ is the energy density of the inflaton, and $m_e$ is electron mass. 
Therefore and because of other dissipation processes, 
the electric fields are dissipated long before their wavelength enters 
the horizon and it could lead to charge separation and anisotropies. 
However, maybe this can lead to fluctuations in the cosmic plasma. 
Hence, 
if the large-scale electric fields with a large amplitude are generated, 
the large-scale charge separation could occur and the large-scale 
fluctuations in the cosmic plasma could be generated. 
Such large-scale charge separation and 
additional fluctuations in the cosmic plasma might have some influence 
on the evolution of the universe. 
For example, if the cosmological-scale charge separation and 
additional fluctuations in the cosmic plasma are generated, the 
evolution of the universe might become anisotropic. This is 
inconsistent with the observational results of the very small anisotropy 
of the CMB radiation obtained from Wilkinson microwave background probe 
(WMAP)~\cite{Spergel03}. 
In this scenario, however, when the large-scale magnetic fields with a 
sufficiently large amplitude are generated during inflation, 
the amplitude of the generated large-scale electric fields is very small. 
Hence the cosmological-scale charge separation and additional 
fluctuations in the cosmic plasma could be hardly generated. Thus, 
this generation scenario of large-scale magnetic fields from inflation is 
consistent with the standard evolution of the universe suggested by the 
observation of CMB radiation. 
This point is the second cosmological significance of the present result.

Finally, we note the following point. 
In the present paper we do not consider a specific model, but we give 
a model-independent analysis on interrelation between the generation of 
large-scale electric and magnetic fields from inflation 
due to the violation of the conformal invariance through the coupling 
$IF^2$. Therefore, we consider the general case in which $I$ is an arbitrary 
function of non-trivial background fields, e.g., the dilaton field $\Phi$ 
and/or the scalar curvature $R$.  
For example, the present author and Yokoyama studied the case of 
the dilaton electromagnetism in Ref.~\cite{Bamba1}, 
where the electromagnetic coupling function is 
given by $I(\Phi)=e^{\lambda\kappa\Phi}$, 
where $\kappa=\sqrt{8\pi G}$ and $G$ is the gravitational constant. 
Such coupling is reasonable in light of indications in higher-dimensional 
theories including string theory. Furthermore, the dilaton potential is given 
by $V[\Phi] = \bar{V} \exp(-\tilde{\lambda} \kappa \Phi)$, 
where $\bar{V}$ is a constant. 
Here, $\lambda$ and $\tilde{\lambda}$ are dimensionless constants. 
The value of the electromagnetic coupling $I(\Phi)$ varies as the 
dilaton $\Phi$ evolves. 
The coupling was first analyzed by Ratra in Ref.~\cite{Ratra}, where 
the inflaton and the dilaton are identified. 
In Ref.~\cite{Bamba1}, the present author and Yokoyama considered a realistic 
situation that the dilaton evolves along with the exponential potential 
$V[\Phi]$ during inflation and even after reheating but is finally stabilized 
when it feels other contributions to its potential, say, from gaugino 
condensation~\cite{GC} that generates a potential 
minimum \cite{Seto, Barreiro}. 
Then they considered the case in which its potential minimum is located 
at $\Phi=0$ and $I=1$ there, and hence the ordinary Maxwell theory is 
recovered. 
As it reaches there, the dilaton starts oscillation with mass $m$ and 
finally decays into radiation with or without significant entropy production.  
As a result, they have shown that if the generated magnetic fields have nearly
scale-invariant spectrum, i.e., $\alpha \approx 4$, 
we had to introduce a huge hierarchy between 
the coupling constant of the dilaton to the electromagnetic field $\lambda$ 
and the coupling one $\tilde\lambda$ of the dilaton potential, 
$\lambda/\tilde{\lambda} \approx 400$.  
As a possible solution to such a huge 
hierarchy between $\lambda$ and $\tilde{\lambda}$,  
the present author and Yokoyama proposed a new scenario~\cite{Bamba2}
by taking account of the effects of 
the stringy spacetime uncertainty relation (SSUR)~\cite{Yoneya}. 
They found that the SSUR of the metric perturbations could lead 
to magnetic fields with a nearly scale-invariant spectrum 
even if $\lambda$ and $\tilde\lambda$ are of the same order of magnitude. 

\section{Conclusion}

In the present paper we have considered the interrelation between 
the generation of large-scale electric fields and that of large-scale 
magnetic fields due to the breaking of the conformal invariance of the 
electromagnetic field through the coupling $IF^2$ in inflationary cosmology. 
As a result, we have shown that if large-scale magnetic fields with a 
sufficiently large amplitude are generated during inflation, the generation of 
large-scale electric fields is suppressed, and vice versa. 
Hence, in this scenario, large-scale magnetic fields with a sufficiently 
large amplitude can be generated during inflation without being inconsistent 
with the fact that the sum of the energy density of the resultant electric and 
magnetic fields during inflation should be smaller than that of the inflaton. 
Furthermore, when large-scale magnetic fields with a 
sufficiently large amplitude are generated during inflation, 
the amplitude of the large-scale electric fields generated is very small. 
Hence the large-scale charge separation and additional fluctuations in the 
cosmic plasma, which could be generated during reheating 
due to the large-scale electric fields and which might 
make the evolution of the universe anisotropic, 
can be hardly generated. 
Consequently, this generation scenario of large-scale magnetic fields from 
inflation is consistent with the standard evolution of the universe 
suggested from the observation of CMB radiation. 
These two points constitute the cosmological significance of the result of 
the present paper.

\section*{Acknowledgements}
This work was supported in part by the 
open research center project at Kinki University.


\end{document}